\documentstyle[prb,aps,multicol,epsf]{revtex}
\tighten
\begin{document}
\draft

\title{Transition from sub-Poissonian to super-Poissonian shot noise
in resonant quantum wells}
\author{Ya. M. Blanter and M. B\"uttiker}
\address{D\'epartement de Physique Th\'eorique, Universit\'e 
de Gen\`eve, CH-1211, Gen\`eve 4, Switzerland} 
\date{\today}
\maketitle

\begin{abstract}
We investigate the transition from
sub-Poissonian to super-Poissonian values
of the zero-temperature shot noise power
of a resonant double barrier of macroscopic cross-section. 
This transition occurs for
driving voltages which are sufficiently strong to bring the system
near an instability threshold. It is shown that interactions in
combination with the energy dependence of the tunneling rates
dramatically affect the noise level in such a
system. Interaction-induced fluctuations of the band bottom of the
well contribute to the noise and lead to a new energy in the Fano
factor. They can enhance the noise to super-Poissonian values in a
voltage range preceding the instability threshold of the system. At
low voltages, interactions may either enhance or suppress the noise
compared to the the non-interacting case.     
\end{abstract}


\begin{multicols}{2}

\section{Introduction}
The investigation of the current noise properties provides 
an important tool to gain additional information on electric systems.
A standard measure of the noise is the Fano factor $F$ which 
is defined as the ratio of the current noise spectral density 
$S_I(\omega)$ and the full Poisson noise 
$2e\langle I \rangle$
where $ \langle I \rangle $ is the average current
driven through the well. Thus the Fano factor 
$F = S_I(\omega)/2e \langle I \rangle$ is smaller than one 
for sub-Poissonian noise and exceeds one for super-Poissonian noise. 
In this work we present theoretical results for the Fano factor
of a resonant double barrier with the macroscopic cross-section and
investigate the transition  from {\it sub}-Poissonian noise to {\it
super}-Poissonian noise.  

An early experiment of Li {\em et al}\cite{Li} focused on the 
suppression of the noise power below the Poisson value. 
Already in this work it was noticed that the Fano factor 
increases from the suppressed values in the voltage range 
of positive differential conductance (PDC) 
as the range of negative differential conductance (NDC)
is approached. Experiments by Brown\cite{Brown} and 
Iannaccone {\em et al}\cite{Iannaccone} showed that in the NDC 
range, the current noise increases even above the Poisson value,
and is thus described by a Fano factor exceeding one. 
Recently, Kuznetsov {\em et al}\cite{Mendez} have presented a detailed
investigation of the noise oscillations from sub-Poissonian to 
super-Poissonian values of a resonant quantum well 
in the presence of a parallel magnetic field. The magnetic field leads 
to multiple voltage ranges of NDC and permits a clear demonstration
of the effect. Below we present a theory which describes 
the transition from sub-Poissonian $(F < 1)$ to super-Poissonian 
noise $(F > 1)$.

Our approach treats the fluctuations of the potential and the
charge in the well. The theory describes the transition from
sub-Poissonian noise to super-Poissonian noise at large voltages. At
smaller voltages it gives either an additional suppression of the
noise power or an enhancement of the noise power compared to theories
which neglect the potential fluctuations. We find that a new quantity,
an energy $\Lambda$, enters the Fano factor. The energy $\Lambda$ is a
measure of the sensitivity of the noise to a collective charge and
current response generated by the Coulomb interaction. The interaction
energy $\Lambda$ is determined by the response of the average current
$\langle I \rangle$ to a variation of the electrostatic potential $U$
in the well   
\begin{eqnarray} \label{ju}
J \equiv \frac{\partial \langle I \rangle}{\partial U}
\end{eqnarray}
and by the response of the charge 
\begin{eqnarray} \label{qu} 
C_0 \equiv
- \frac{\partial \langle Q \rangle}{\partial U}
\end{eqnarray}
to the potential in the well. $C_{0}$ has the dimension of
capacitance. It is positive in the PDC range, giving rise to screening
of excess charges, and becomes negative as the NDC range is approached,
corresponding to {\em overscreening} of excess charges. For a double
barrier with a geometrical capacitance $C_L$ across the left barrier
and a geometrical capacitance $C_R$ across the right barrier we find
an interaction energy  
\begin{eqnarray} \label{Fu} 
\Lambda \equiv \frac{\hbar {J}}{C_L + C_R + C_0}. 
\end{eqnarray} 
Note that $J$ has the dimension of a conductance. The interaction 
energy thus has the form of a dimensionless conductance 
$\hbar J/e^{2}$ multiplied by an effective charging energy 
of the well $e^{2}/(C_L + C_R + C_0)$. In the NDC range
the charging energy grows and leads to a very large 
interaction energy $\Lambda$. 
For a resonant level characterized by decay widths $\Gamma_{L,R}$
through the left and right barrier,
an asymmetry $\Delta \Gamma = (\Gamma_{L} - \Gamma_{R})/2$, and 
a total width $\Gamma = \Gamma_{L} + \Gamma_{R}$, 
we find a Fano factor given by 
\begin{eqnarray} \label{Fano1}
F = \frac{1}{2} + 2 
\frac{(\Lambda - \Delta \Gamma)^{2}}{\Gamma^2} . 
\end{eqnarray}
Note that the escape rates $\Gamma_{L,R}$ are positive quantities
but $\Lambda$ and $\Delta \Gamma $ can have either sign. 
For a symmetric double barrier the sign of $\Delta \Gamma$ is
independent of the applied voltage. However, as will be shown in this
work, for any barrier $\Lambda$ changes sign as a function of voltage:
It is negative in the stable part of the I-V-characteristic and
becomes positive and large in the NDC part of the I-V
characteristic. 

If $\Lambda$ is set to zero in Eq. (\ref{Fano1}), the Fano factor 
for non-interacting electrons of Chen and Ting\cite{ChenTing} is
recovered. This result is a consequence of partition noise: an
electron scattered at an obstacle with transmission probability T can
only be either transmitted or reflected. The partition
noise\cite{Khlus,B90,BdJ} gives rise to shot noise proportional to
$T(1-T)$. For a resonant tunneling barrier\cite{B91}, 
using the Breit-Wigner expression for the transmission 
probability, this  leads immediately to the Fano factor,
Eq. (\ref{Fano1}) with $\Lambda = 0$.
For $\Lambda = 0$ the 
Fano factor is minimal (equal to $F = 1/2$) if the two barriers 
lead to equal escape rates $\Delta \Gamma = 0$
and the Fano factor is maximal (with a value of $F = 1$),  if 
the double barrier structure is very asymmetric,  $|\Delta \Gamma |
\simeq \Gamma /2 $. The lower limit $F = 1/2$ is obtained, independent
of whether transmission is a quantum mechanical, fully coherent
process or a sequential two step process without phase
coherence\cite{Chen2,Davies1,Levy,Davies3,Iannaccone1}. This maximal
suppression is also obtained in the presence of a magnetic field (see
Ref. \onlinecite{Bo}). On the other hand, models with internal degrees
of freedom\cite{Davies2} can give rise to a noise suppression slightly
below $F = 1/2,$ if the internal degrees of freedom couple also to the
reservoirs.   
 
Our result, Eq. (\ref{Fano1}), also has a Fano factor of 
$1/2$ as a lower limit. However, depending on the asymmetry, 
at low voltages our theory predicts either a reduction 
or enhancement of the Fano factor compared to a free-electron theory. 
A negative $\Lambda$ arises if an increase of the internal 
voltage $U$ decreases the current through the well. 
This is the case in the stable range of the I-V characteristic:
a charging of the well is counteracted  
by the Coulomb interaction which creates a positive potential 
response which in turn drives the resonant level 
further up in energy. This leads to a Coulomb induced 
correction of the Fano factor for a resonant double well  
even at low voltages. 

As the range of the NDC is approached, the interaction energy
$\Lambda$ changes sign and becomes positive to such a degree that it
completely dominates the Fano factor. This large increase is again due
to the Coulomb interaction.  However, now we are in a range where the
average charge increases in response to a positive variation of the
internal voltage. A charge injected into the well creates a
fluctuation of the internal potential which causes even more charge to
flow into the well. Thus the current fluctuations are magnified and
give rise to a Fano factor which exceeds the Poisson value.   

Large fluctuations are commonly expected near equilibrium 
or non-equilibrium phase transitions. The analogy of the far from
equilibrium current instabilities with equilibrium phase transitions
was the subject of work by Pytte and Thomas\cite{PT}. These authors
investigated the critical slowing down of dielectric relaxation modes
at a bulk current instability (Gunn effect). Subsequent work by
B\"uttiker and Thomas\cite{BT} analyzed a one band-current instability
(superlattice) and discussed in addition to the  dielectric relaxation
modes also the modes which arise due to the coupling of the current
fluctuations to the magnetic field. In this work we investigate the
resonant well under the condition of complete external voltage control
and take into account the electrostatic but not the magnetic part of
the Coulomb interaction.  

There are two theoretical works\cite{Brown,Anwar}
which also indicate Fano factors in excess of the Poisson value. 
However, these works do not include the partition noise. 
Instead, it is assumed that 
the incident current on the cathode side is fully Poissonian 
but that the transmitted and reflected carrier streams are modified
through the current dependence of the 
transmission probabilities. Our theory 
takes as the basic noise source the partition noise and treats
the associated charge fluctuations self-consistently.
To treat the potential fluctuations,
we evaluate the off-diagonal elements of a potential operator
of the well. In contrast, Ref. \onlinecite{Brown} does not treat
potential fluctuations and Ref. \onlinecite{Anwar} includes
self-consistent effects only on the level of probabilities. These
works\cite{Brown,Anwar} also predict in the PDC range Fano factors
lower than $F = 1/2$. Such low Fano factors have been observed in the
recent experiment by Kuznetsov {\em et al}\cite{Mendez}. However,
since in Refs. \onlinecite{Brown,Anwar} an essential source of noise,
the partition noise, which gives raise to the minimal Fano factor $F
=1/2$ in the non-interacting case, is absent, it is perhaps not
surprising that lower noise limits are predicted than by our theory.

Below we present a self-consistent, analytical calculation of
transport properties of resonant tunneling quantum wells in the limit
when the total decay width of the resonant state is lower than all
other energy scales of the problem. We consider the zero-temperature
case and use the system of units with $\hbar = 1$. 
We continue to use the expression "shot noise" 
even in the case when the noise is dominated by interaction.
We emphasize, that in this case, we deal not with single independent
electrons traversing the conductor, but with a collective response
of many electrons. The term ``shot noise'' underlines here that 
the basic source of noise is the partition noise, 
independent of the degree of interaction.    

The paper is organized as follows: In Section II we introduce the
self-consistency equation and solve it for the average
quantities. Sec. III is devoted to the average current. The results of
these two sections by themselves are not new, and similar discussion
can be found in the literature (see {\em e.g.}
Ref. \onlinecite{Levy}). However, they are necessary
for the subsequent calculation of the current-current fluctuations.
This calculation is performed in Sec. IV. In Sec. V we present the
conclusions and discuss in more detail the conditions of validity of
our theory.   

\section{Average charge and potential} 

\subsection{The charge of the well and the self-consistency equation}

We consider a quantum well\cite{Tsu} extended in the $x$ and
$y$-directions (to be denoted as $\perp$) with the area $\cal{A}$. The
potential profile in the $z$-direction is shown in Fig.~1. We assume
that the longitudinal and the transverse motion of the electrons are
separable. The motion in the $z$-direction is quantized, and there
exists a set of resonant levels with energies $E_n$,
$n = 0,1, \dots$. The transverse motion is described by a continuum
of momenta $p_{\perp}$.  Thus the expression for the total energy
(measured from the band bottom of the well) is 
$$E_{n,p_{\perp}} = E_n + p_{\perp}^2/2m.$$
The density of states in the well can be factorized into longitudinal
and transverse parts,
$$\nu(E) = \nu_2 \nu_z (E_z), \ \ \ E = E_z + p^2_{\perp}/2m.$$
Here  $\nu_2 = m/2 \pi$ is the
the two-dimensional density of states (per spin). 
\begin{figure}
\narrowtext
{\epsfxsize=5.0cm\epsfysize=5.0cm\centerline{\epsfbox{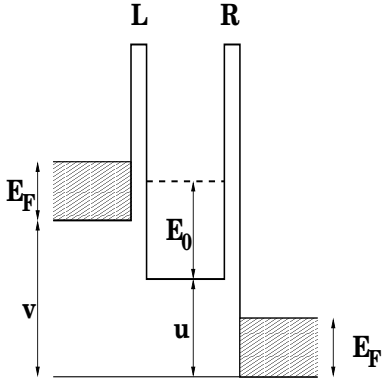}}}

\protect\vspace{0.5cm}

\caption{The potential profile of the quantum well.} 
\label{fig1}
\end{figure}
\noindent

For interacting electrons the band bottom of the well $u=eU$ 
(see Fig. (\ref{fig1})) is not an independent quantity but must be
found from a self-consistency equation. In the present work, we
account for the interaction by including the energy required 
to charge the well. We introduce
the capacitances $C_L$ and $C_R$ of the left and the right barrier,
respectively. The charge allowed in the well by the Coulomb interaction 
is then determined by these capacitances and the voltage drop 
across the left barrier $u-v$ and across the right barrier $u$. 
Here $v>0$ is the voltage applied across the double barrier structure.
(We assume that the Fermi energies of the left and right contact are
equal). Thus the total charge $Q[u]$ in the well is 
\begin{equation} \label{self-cons}
C_L (u-v) + C_R u = eQ[u]. 
\end{equation}
Note that the charge $Q$ is always non-negative, and thus the band
bottom either assumes the value 
$u_{e} = (C_L+C_R)^{-1}C_Lv$ of the empty well, 
or lies above it. 

Throughout this work, we describe scattering
at the double barrier in the Breit-Wigner limit. In
particular, the longitudinal density of states is 
\begin{equation} \label{DOS}
\nu_z(E_z) = \frac{1}{2\pi} \sum_n \frac{\Gamma(E_z)}{(E_z - E_n)^2
+ \Gamma^2(E_z)/4}.
\end{equation}
Here we have introduced the total width of a single resonance
$\Gamma(E_z) = \Gamma_L(E_z) + \Gamma_R(E_z)$ with $\Gamma_L(E_z)$ and
$\Gamma_R(E_z)$ being the partial widths for decay 
through the left and right barrier, respectively. 
The Breit-Wigner limit is appropriate in the limit of small
tunneling probabilities through the two barriers
and provides an excellent description as long as 
the resonant level is not 
in the immediate vicinity of the band bottom of any reservoir,
$u + E_n  = v$.     

In the standard application of the Breit-Wigner expression
the partial width is independent of energy; for energies 
$E_z$ close to the resonant state with energy $E_n$ the 
width is evaluated at the energy of the resonant state,
$\Gamma(E_n)$. Here, in order to describe the density of states in the
limit where the resonant level approaches the conduction band bottom,
we consider the decay widths to be energy dependent. 
We determine the energy dependence by considering the 
quantum-mechanical problem of transmission
through a single rectangular barrier. 
We write   
\begin{eqnarray} \label{rates}
\Gamma_L (E_z) & = & a_L E_z^{1/2} (E_z + u - v)^{1/2} \theta(E_z)
\theta(E_z + u - v); \nonumber \\
\Gamma_R (E_z) & = & a_R E_z^{1/2} (E_z + u)^{1/2} \theta(E_z)
\theta(E_z + u), 
\end{eqnarray}
where $a_L$ and $a_R$ are dimensionless parameters. 

With the density of states as specified above, we can now 
find the charge $Q[u]$ inside the well. 
For this purpose, we divide the total longitudinal
density of states $\nu_z(E)$ given by Eq. (\ref{DOS}) into the 
partial densities of states of
electrons which are incident from the left or the right reservoirs,
$$\nu_z(E_z) = \nu_L (E_z) + \nu_R(E_z).$$
We find\cite{Christen} 
\begin{equation} \label{inject}
\nu_{L,R} (E_z) = \frac{\Gamma_{L,R} (E_z)}{\Gamma(E_z)} \nu_z(E_z).
\end{equation}
With the distribution functions $f_{L,R}$ of the left and right
reservoir and the short-hand notation $E_{\perp} \equiv
p_{\perp}^2/2m$, we find for the charge of the well
\begin{equation} \label{charge1}
Q[u] = e \sum_{p_{\perp}} \int dE_z \sum_{i = L,R}
\nu_i (E_z) f_i (E_z + E_{\perp}).  
\end{equation}
Eq. (\ref{charge1}) gives the total charge in the well 
in terms of the carriers injected from each reservoir. 

To proceed we now specify two additional conditions: 
\begin{description}
\item{(i)} Of all longitudinal energy levels only $E_0$ is relevant,
all others lie too high. In addition, we assume for definiteness $E_0
> E_F$. 

\item{(ii)} The decay widths $\Gamma_L$ and $\Gamma_R$ are much
smaller then all other energy scales. In particular, we will replace
$(\xi^2 + \Gamma^2/4)^{-1}$ by $2\pi \Gamma^{-1} \delta(\xi)$, where
appropriate. 
\end{description}

Making use of these conditions and replacing the sum over $p_{\perp}$
by an integral, we obtain
\end{multicols}
\widetext
\vspace*{-0.2truein} \noindent \hrulefill \hspace*{3.6truein}
\begin{equation} \label{charge2}
Q[u] = \left\{ \matrix{
C_e \displaystyle{\Gamma_L(E_0)\over{\Gamma(E_0)}} (v +
E_F - u - E_0), 
& v - E_0 < u < v - E_0 + E_F \cr
0, & \mbox{\rm otherwise .}
} \right. ,
\end{equation} 
\hspace*{3.6truein}\noindent \hrulefill 
\begin{multicols}{2}
\noindent
Here we have introduced the ``quantum capacitance'' 
$C_e \equiv e\nu_2 {\cal{A}}$ of the two-dimensional 
electron gas. 
The dependence of $Q[u]$ on $u$ is shown
schematically in Fig.~2 (solid line).  

\subsection{Self-consistent calculation of the band bottom $u$}

Equation (\ref{self-cons}) expresses the band bottom in the
well $u$ as a function of the applied voltage $v$. The equation is
cubic and can be solved analytically. However, the resulting
expressions are cumbersome and not very transparent. Therefore, we
discuss now the behavior of the function $u(v)$ qualitatively, and
support the discussion by numerical results (Fig.~2).

The left-hand side of Eq. (\ref{self-cons}) turns to zero 
(empty well) for $u = (C_L +C_R)^{-1}C_L v$. 
The charge is non-zero only between $u =
v-E_0$ and $u = v - E_0 + E_F$. With the help of the voltages 
$$v_a \equiv \frac{C_L + C_R}{C_R}\left( E_0 - E_F \right); \ \ \ v_b
\equiv \frac{C_L + C_R}{C_R}E_0,$$
we can distinguish the following four regimes: 

\begin{description}

\item{I}.   $0 < v < v_a$ (Fig.~2a). The well is uncharged; the only
solution of Eq. (\ref{self-cons}) is,
\begin{equation} \label{sol1}
u(v) = u_e \equiv \frac{C_L}{C_L + C_R}v.
\end{equation} 

\item{II}.  $v_a < v < v_b$ (Fig.~2b). The well is now charged, but
Eq. (\ref{self-cons}) has still only one solution.

\item{III}. $v_b < v < v^*$ (Fig.~2c). For voltages above $v_b$ three
solutions emerge; now $u_1(v)$ is given by Eq. (\ref{sol1}) and
corresponds to the uncharged well, while two others $u_2(v)$ and 
$u_3(v)$ describe a charged well. The lesser of them, $u_2(v)$, is
unstable, and the greater one, $u_3(v)$, is stable. At a certain
voltage $v^*$, which is a complicated function of the parameters of
the system, the straight line $e^{-1}[C_L(u-v) + C_Ru]$ is a tangent
to $Q[u]$. This voltage determines the upper margin for this regime.

\item{IV}. $v^* < v$ (Fig.~2d). The well is again uncharged, and the
only solution is given by Eq. (\ref{sol1}).

\end{description}

The dependence $u[v]$ is illustrated in Fig.~3. 

\section{Average current}

To find the current across the quantum well, we start from the
general expression for the current density,
\begin{eqnarray} \label{cur1}
j & = & \frac{e}{2\pi{\cal{A}}} \sum_{p_{\perp}} \int dE_z T(E_z)
\nonumber \\
& \times & \left[
f_L(E_z + E_{\perp}) - f_R(E_z + E_{\perp}) \right], 
\end{eqnarray}
where $T(E_z)$ is the transmission coefficient, equal to
\cite{doublebar} 
\begin{equation} \label{transm}
T(E_z) = \frac{\Gamma_L(E_z)\Gamma_R(E_z)}{(E_z - E_0)^2 +
\Gamma^2(E_z)/4} .
\end{equation}
Treating Eq. (\ref{cur1}) in the same way as previously
Eq. (\ref{charge1}), we obtain
\begin{equation} \label{cur2}
j = \frac{C_e}{{\cal{A}}}
\frac{\Gamma_L(E_0)\Gamma_R(E_0)}{\Gamma(E_0)} (v + E_F - u - E_0) 
\end{equation}
for $v < u + E_0 < v + E_F$ and zero otherwise.  

This can be easily translated into the current-voltage
characteristics, given the known dependence $u(v)$. The current is
zero for $v < v_a$ and $v > v^*$. For $v_a < v < v_b$ it is given by
Eq. (\ref{cur2}), where $u$ is replaced by the only solution $u(v)$.
Finally, for $v_b < v < v^*$ the three possible solutions
$u(v)$ yield three values for the current for any voltage. One of them
is $j = 0$ and corresponds to the uncharged well (the solution
$u_1(v)$). Two other branches correspond to the charged well; the
stable branch (3 in Fig. 3, due to $u_3(v)$) lies above the unstable
one (2). The current-voltage characteristics is shown in Fig.~4 (solid
line). The differential resistance jumps from zero to a finite
positive value at $v = v_a$ and then decreases with voltage, passing
through zero at a certain point. It is negative for voltages close to
$v^*$ and turns to $-\infty$ at this point.  

In the experiment the system is expected to exhibit a hysteretic
behavior (cf. Ref. \onlinecite{Levy}). Upon increasing the voltage,
the current follows the upper branch (3 in Fig.~4) until $v = v^*$,
and then jumps onto the zero-current branch. Upon decreasing the
voltage, the current is zero until $v = v_b$, and then jumps to a
finite value on the branch 3. The unstable branch 2 is not
observed. Note that the range of hysteresis $v^* - v_b$ is broader
than the range of negative differential resistance.  
\begin{figure}
\narrowtext
{\epsfxsize=5.0cm\epsfysize=5.0cm\centerline{\epsfbox{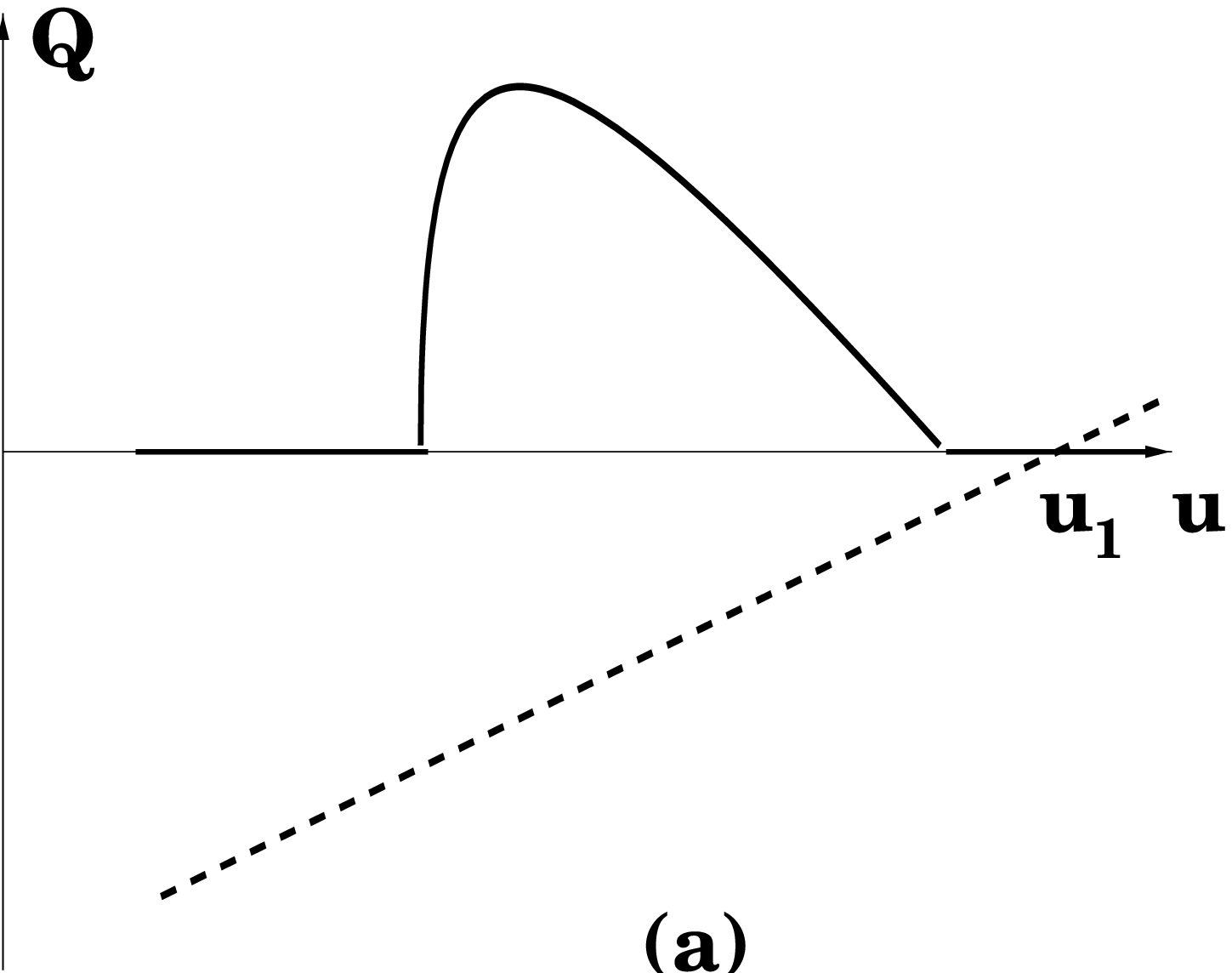}}}
{\epsfxsize=5.0cm\epsfysize=5.0cm\centerline{\epsfbox{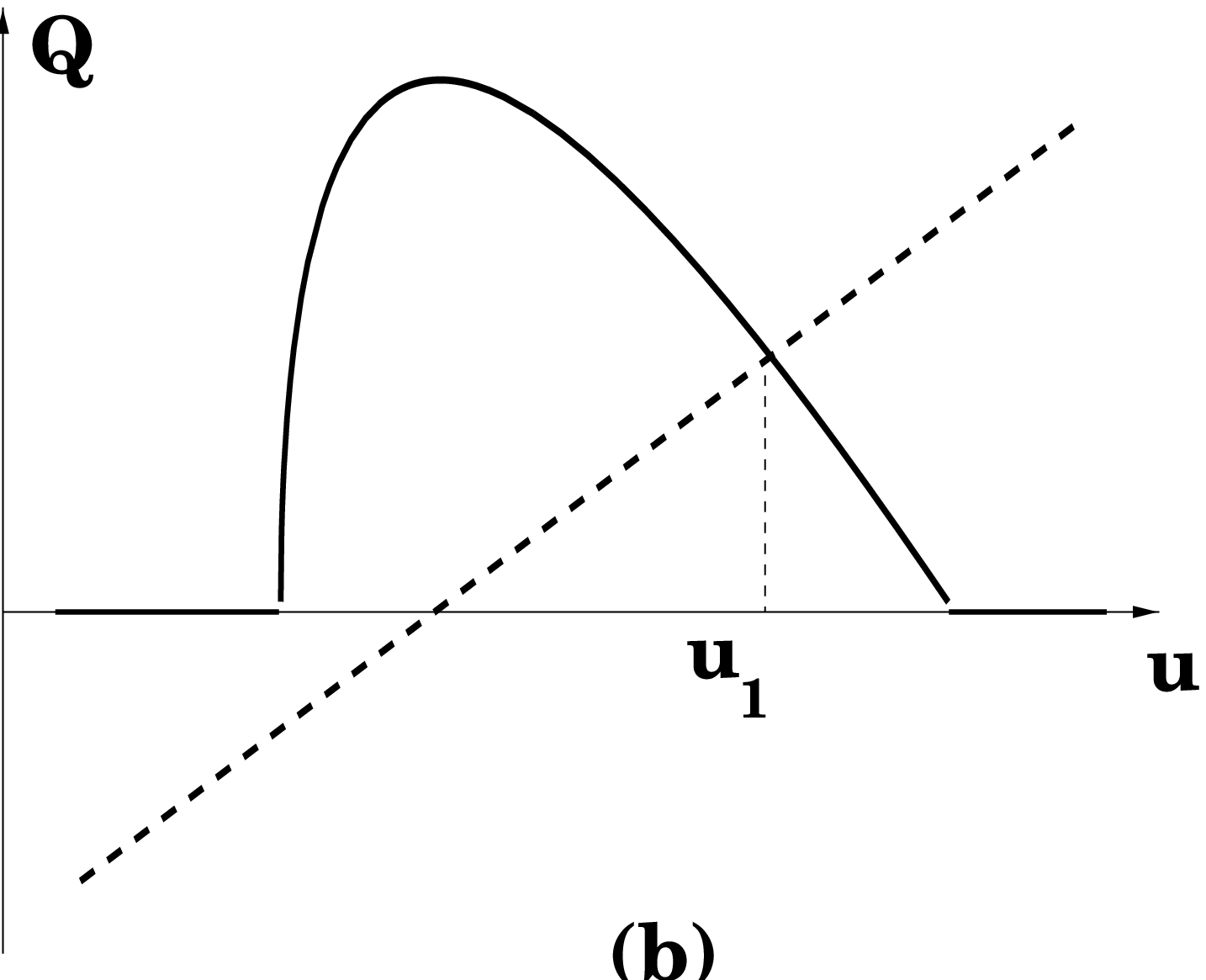}}}
{\epsfxsize=5.0cm\epsfysize=5.0cm\centerline{\epsfbox{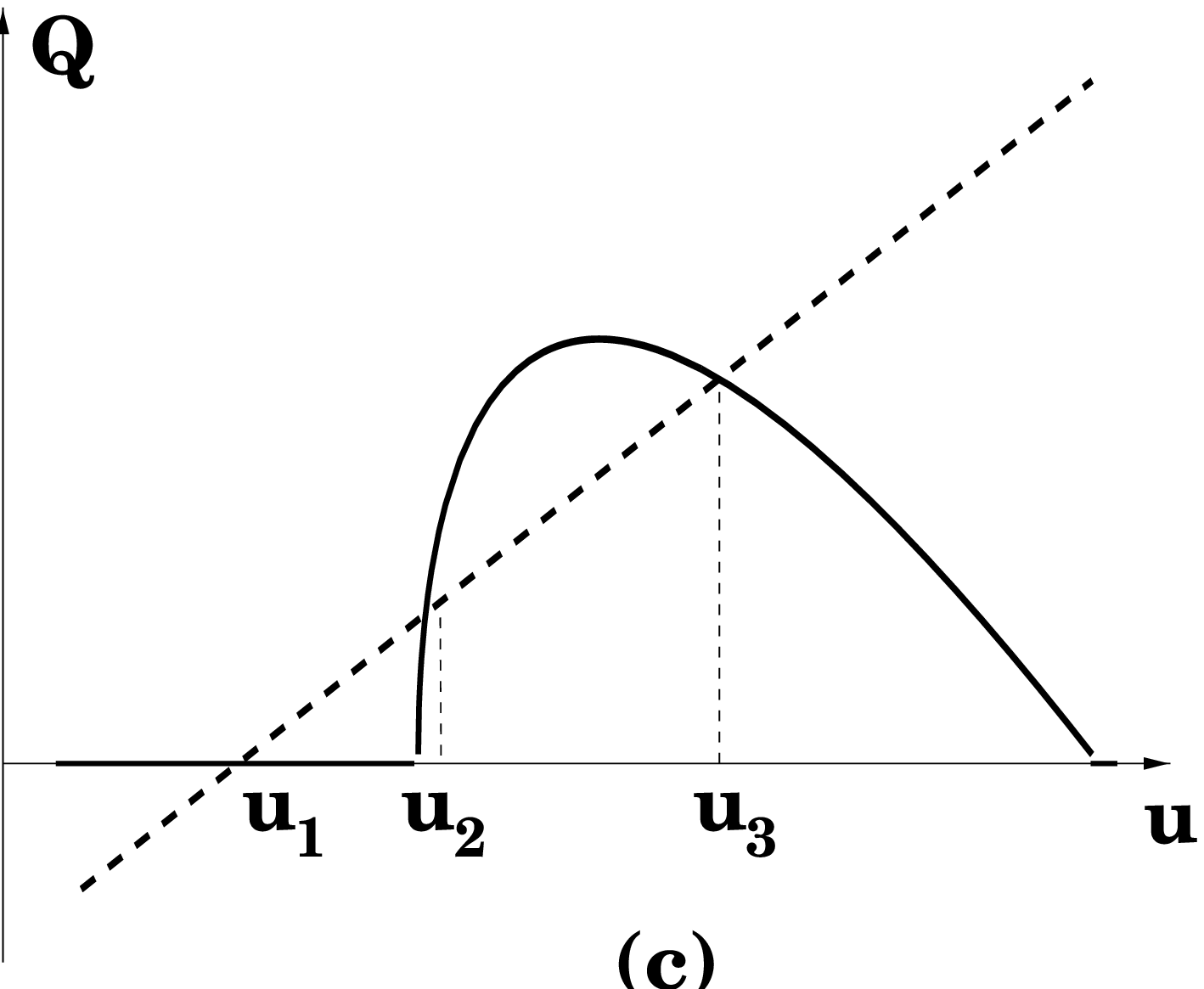}}}
{\epsfxsize=5.0cm\epsfysize=5.0cm\centerline{\epsfbox{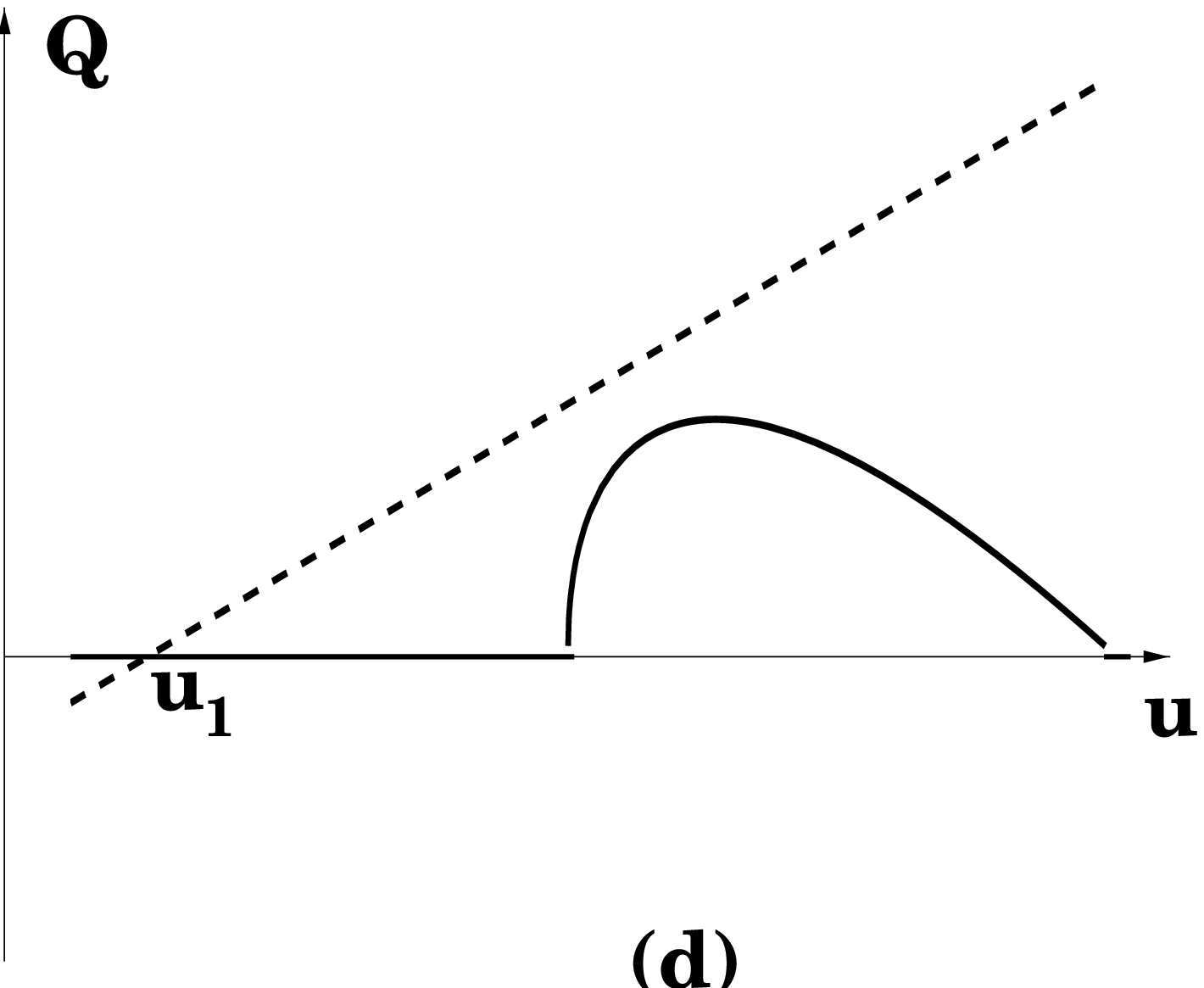}}}

\protect\vspace{0.5cm}

\caption{Solutions of the self-consistency equation
(\protect\ref{self-cons}). The solid line represents the charge
$Q[u]$; the dashed line is $e^{-1}[C_L(u-v) + C_Ru]$. The position of
the vertical axis is arbitrary. The following set of parameters is
chosen: $a_L = a_R$; $C_L = C_R = C_e/10$; $E_0 = 3E_F/2$. Voltages
are (in units of $E_0$) 1/3 (a), 1 (b), 7/3 (c), 3 (d). The voltage
$v^*$ is equal to 2.70. The solutions are shown as $u_i$; in the case
(c) there are three solutions, of which $u_1$ and $u_3$ are stable,
and $u_2$ is unstable.}   
\label{fig2}
\end{figure}
\noindent

The above analysis is valid in the case of low transparency,
$\Gamma(E_0) \ll v^* - v_b$. In the opposite case the hysteretic range
is smeared, and a broad range of negative differential conductance 
appears instead. For weak interaction $C_e \ll C_L, C_R$ the width of
the hysteretic range is rather small,
$$v^* - v_b  = \left( \frac{a_L}{2a_R} \frac{C_e}{C_L + C_R}
\right)^2 \frac{E_F^2}{E_0},$$  
but it increases with the capacitance $C_e$ and eventually must exceed
the smearing $\Gamma(E_0)$ of the resonant level. 
\begin{figure}
\narrowtext
{\epsfxsize=5.0cm\epsfysize=5.0cm\centerline{\epsfbox{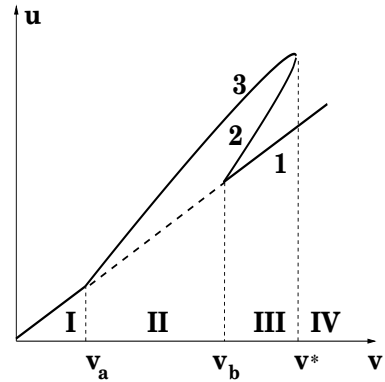}}}

\protect\vspace{0.5cm}

\caption{Dependence $u(v)$ (solid line). Curves 1, 2, and 3 in the
regime III correspond to the solutions $u_1$, $u_2$, and $u_3$,
respectively. The solution $u_2$ is unstable. The parameter set is the
same as for Fig.~2.} 
\label{fig3}
\end{figure}
\noindent

\section{Fluctuations}

\subsection{Charge operator}

The fluctuations of the charge, the potential and the current are 
determined by the off-diagonal elements of these quantities.  
We must use operator expressions instead of
average quantities. Following Ref. \onlinecite{Buttiker92}, we
introduce creation and annihilation operators $\hat
a^{\dagger}_{\alpha} (E_z,p_{\perp})$ and $\hat a_{\alpha}
(E_z,p_{\perp})$ of carriers in the reservoir $\alpha = L,R$. The
quantum statistical average of bilinear products of these operators is
\begin{eqnarray} \label{av1}
& & \left\langle \hat a^{\dagger}_{\alpha} (E_z,p_{\perp}) \hat
a_{\beta} (E'_z,p'_{\perp}) \right\rangle = \delta_{\alpha\beta}
\delta_{p_{\perp} p'_{\perp}} \nonumber \\
& & \times \delta(E_z - E'_z) f_{\alpha} (E_z + p_{\perp}^2/2m).
\end{eqnarray}

Here we are interested in the zero-frequency spectral correlations. In
this limit the off-diagonal elements of the charge operator
can be obtained with the help of off-diagonal density of 
states elements which in turn can be related to energy 
derivatives of the scattering
matrix. In the presence of a current amplitude incident 
at contact $\alpha$ and a current amplitude at contact $\beta$
the off-diagonal density of states element 
$\nu_{\alpha\beta}$,  
in terms of scattering matrices $s_{\alpha\beta}$, is
\cite{Buttiker96,Pedersen} 
\begin{equation} \label{partdens}
\nu_{\alpha\beta} (E_z) = \frac{1}{4\pi i} \sum_{\gamma} \left[
s^{\dagger}_{\gamma\alpha} \frac{ds_{\gamma\beta}}{dE_z} -
\frac{ds^{\dagger}_{\gamma\alpha}}{dE_z} s_{\gamma\beta} \right].
\end{equation}
Using the explicit expressions for the scattering matrices
\cite{Christen}, 
\begin{equation} \label{scatmatr}
s_{\alpha\beta} = \exp \left[i \left(\phi_{\alpha} + \phi_{\beta}
\right) \right] \left( \delta_{\alpha\beta} -
i\frac{\sqrt{\Gamma_{\alpha}\Gamma_{\beta}}}{E_z - E_0 + i\Gamma/2}
\right)
\end{equation}
(with $\phi_{\alpha}$ being arbitrary phases), and having in mind that
the main singularity is generated by the derivative of the denominator
in Eq. (\ref{scatmatr}), we easily obtain
\begin{equation} \label{part2}
\nu_{\alpha\beta} (E_z) = \frac{1}{2\pi} \frac{\sqrt{\Gamma_{\alpha}
(E_z) \Gamma_{\beta} (E_z)}}{(E_z - E_0)^2 + \Gamma^2(E_z)/4}
e^{i(\phi_{\beta} - \phi_{\alpha})}.
\end{equation}
With the help of these off-diagonal density of states 
elements the 
operator of the charge of the well is 
\cite{Buttiker96} 
\begin{eqnarray} \label{chargeop1}
\hat Q(t) & = & e\sum_{\alpha\beta} \sum_{p_{\perp}} \int d\omega
e^{- i\omega t} \int dE_z \nu_{\alpha\beta} (E_z) \nonumber \\
& \times & \hat a^{\dagger}_{\alpha}
(E_z,p_{\perp}) \hat a_{\beta} (E_z + \omega, p_{\perp}). 
\end{eqnarray}
It is straightforward to check that Eq. (\ref{chargeop1}) reproduces
the average charge (\ref{charge1}). Below we will use the charge
operator to find the fluctuations. 
\begin{figure}
\narrowtext
{\epsfxsize=5.0cm\epsfysize=5.0cm\centerline{\epsfbox{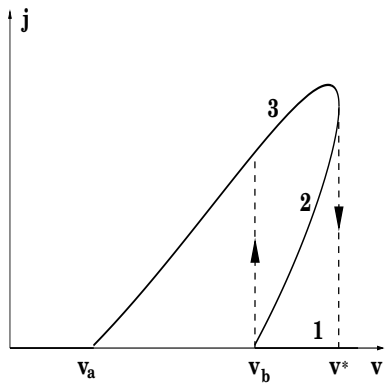}}}

\protect\vspace{0.5cm}

\caption{The current-voltage characteristics $j(v)$ (solid line). The
hysteretic behavior is shown by dashed lines. The parameter set is the
same as for Fig.~2.} 
\label{fig4}
\end{figure}
\noindent

\subsection{Band bottom operator}

With $Q$ being an operator, the self-consistency equation
(\ref{self-cons}) must be considered as an operator equation. Thus, we
are obliged to introduce the band bottom operator $\hat u$, which is a
solution of Eq. (\ref{self-cons}). A similar approach has already been
used in Ref. \onlinecite{Pedersen}. The charge operator $\hat Q$
becomes then a superoperator $\hat Q [\hat u]$. In the following we
consider only the voltage range $v_a < v < v^*$, and in case of
multiple stable states only the one corresponding to the charged well
(denoted $u_3(v)$ in the preceding discussion). 

Eq. (\ref{self-cons}) is non-linear. Instead of trying to find a
general solution, we linearize it, writing $\hat u = u + \delta\hat
u$, where $u$ is the solution of Eq. (\ref{self-cons}) for the average
quantities. With the ``parameter'' of the expansion being
$\hat a^{\dagger}_{\alpha} \hat a_{\beta} - \langle \hat
a^{\dagger}_{\alpha} \hat a_{\beta} \rangle$, we write the charge
operator as follows,
\begin{eqnarray} \label{charge4}
\hat Q [\hat u] & = & \langle Q[u] \rangle + \hat Q_1 + \hat Q_2,
\nonumber \\ 
\hat Q_1 & = & - \frac{C_0[u]}{e} \delta \hat u; \ \ \ C_0[u] \equiv
- e\frac{\partial \langle Q \rangle}{\partial u}; \\  
\hat Q_2 & = & e\sum_{\alpha\beta} \sum_{p_{\perp}} \int d\omega
e^{- i\omega t} \int dE_z \nu_{\alpha\beta} (E_z;u) \nonumber \\
& \times & \left[ \hat
a^{\dagger}_{\alpha} (E_z,p_{\perp}) \hat a_{\beta} (E_z + \omega,
p_{\perp}) - \right. \nonumber \\
& & - \left. \left\langle \hat a^{\dagger}_{\alpha} (E_z,p_{\perp})
\hat a_{\beta} (E_z + \omega, p_{\perp}) \right\rangle
\right]. \nonumber 
\end{eqnarray} 
Here the average charge $\langle Q \rangle$ is given by
Eq. (\ref{charge2}) and is a function of the average 
potential $u$ (not of the operator $\hat u$). The
capacitance\cite{note1} $C_{0}[u]$ determines the charge increment in
the well in response to an increment of the potential in the well. It
is one of the important response functions which determines the new
parameter $\Lambda$ which eventually enters the Fano factor and was,
therefore, mentioned already in the introduction (see Eq. (1), $U =
u/e$). On the stable branch (3 in Fig.~3) the ``capacitance'' $C_0$
changes sign, and for the voltage $v^*$ it is negative and equal to
$-(C_L + C_R)$. The operator $\hat Q_2$ appears as the operator for
the bare (unscreened) charge and is a function of the average
potential $u$. For the operator $\delta \hat u$ we obtain 
\begin{equation} \label{band1}
\delta \hat u (t) = \frac{e}{C_L + C_R + C_0[u]} \hat Q_2(t).
\end{equation} 
For $v \to v^*$ this expression diverges. We will show later that this
divergence is at the origin of the noise enhancement as the voltage
$v^*$ is approached. 

\subsection{Current operator}

Following Ref. \onlinecite{Buttiker92}, we write the current density
operator in the contact $\alpha$, 
\begin{eqnarray} \label{cur3}
\hat j_{\alpha} (t) & = & \frac{e}{2\pi{\cal{A}}} \sum_{\beta\gamma}
\sum_{p_{\perp}} \int dE_z dE'_z e^{i(E_z - E'_z)t}
A^{\alpha}_{\beta\gamma} (E_z,E'_z) \nonumber \\
& & \times \hat a^{\dagger}_{\beta}
(E_z,p_{\perp}) \hat a_{\gamma} (E'_z, p_{\perp}), 
\end{eqnarray}
with current matrix elements 
\begin{equation} \label{coef}
A^{\alpha}_{\beta\gamma} (E_z,E'_z) = \delta_{\alpha\beta}
\delta_{\alpha\gamma}  - s^{\dagger}_{\alpha\beta} (E_z)
s_{\alpha\gamma} (E'_z)
\end{equation}
which are themselves functions of the operator $\hat u$. Using
Eq. (\ref{scatmatr}), we obtain
\begin{eqnarray} \label{coef2}
A^L_{LL} (E_z,E_z) = A^R_{RR} (E_z,E_z) = - A^L_{RR} (E_z, E_z)
\nonumber \\
= -A^R_{LL} (E_z,E_z) = \frac{\Gamma_L(E_z)\Gamma_R(E_z)}{(E_z-E_0)^2
+ \Gamma^2/4}.  
\end{eqnarray}
It is immediately seen that $\langle j_L \rangle = - \langle j_R
\rangle$ and the average current $\langle j_L \rangle$ reproduces
Eq. (\ref{cur1}).  

The current operator (\ref{cur3}) also must be linearized in $\delta
\hat u$ in the same manner as we did it for the charge operator. We
obtain 
\begin{eqnarray} \label{cur4}
\hat j_{\alpha} [\hat u] & = & \langle j_{\alpha} [u] \rangle + \hat
j_{1\alpha} + \hat j_{2\alpha}, 
\nonumber \\ 
\hat j_{1\alpha} & = & J_{\alpha} \delta \hat u; \ \ \ J_{\alpha}
\equiv \frac{\partial \langle j_{\alpha} \rangle}{\partial u}; \\  
\hat j_{2\alpha} & = & \frac{e}{2\pi{\cal{A}}} \sum_{\beta\gamma}
\sum_{p_{\perp}} \int dE_z dE'_z e^{- i(E'_z-E_z) t}
A^{\alpha}_{\beta\gamma} (E_z;E'_z) \nonumber \\
& \times & \left[ \hat
a^{\dagger}_{\beta} (E_z,p_{\perp}) \hat a_{\gamma} (E'_z, p_{\perp})
- \right. \nonumber \\ 
& & - \left. \left\langle \hat
a^{\dagger}_{\beta} (E_z,p_{\perp}) \hat a_{\gamma} (E'_z, p_{\perp})
\right\rangle \right]. \nonumber 
\end{eqnarray} 
The current operator $\hat j_{1\alpha}$ is proportional 
to $J_{\alpha}$, the increment in current at contact $\alpha$ 
in response to a potential variation in the well. 
$J_{\alpha}$ is the second important response function 
which enters the Fano factor and has also been mentioned
in the introduction (see Eq. (2)). $\hat j_{2\alpha}$
is the bare current density and via the current matrix  
$A^{\alpha}_{\beta\gamma}$ depends on $u$ (not on
$\hat u$). Note that $J_L = -J_R$ and that in the introductory section
we have suppressed the contact index and have defined $J \equiv e
{\cal{A}} J_L = -e {\cal{A}} J_R$, $U = u/e$.

\subsection{Current-current fluctuations and the Fano factor}

Now we are in a position to evaluate the shot noise spectral power
$S_{\alpha\beta}$, defined as \cite{Buttiker92}
\end{multicols}
\widetext
\vspace*{-0.2truein} \noindent \hrulefill \hspace*{3.6truein}
\begin{eqnarray} \label{noise1}
2\pi \delta(\omega + \omega') S_{\alpha\beta}(\omega) = 
\int dt dt' e^{i\omega t + i \omega t'} 
\left[ \left\langle \hat
j_{\alpha} (t) \hat j_{\beta} (t') + \hat j_{\beta} (t') \hat
j_{\alpha} (t) \right\rangle - 2\left\langle \hat j_{\alpha} (t)
\right\rangle \left\langle \hat j_{\beta} (t') \right\rangle \right]. 
\end{eqnarray}
We consider the zero-frequency
limit and use the property \cite{Buttiker92}
\begin{eqnarray} \label{aver4}
\left\langle \hat a^{\dagger}_{\alpha} (E_1,p_{\perp}) \hat a_{\beta}
(E_2,p_{\perp}) \hat a^{\dagger}_{\gamma} (E_3,p'_{\perp}) \hat
a_{\delta} (E_4, p'_{\perp}) \right\rangle - \left\langle \hat
a^{\dagger}_{\alpha} (E_1,p_{\perp}) \hat a_{\beta} 
(E_2,p_{\perp}) \right\rangle \left\langle \hat a^{\dagger}_{\gamma}
(E_3,p'_{\perp}) \hat a_{\delta} (E_4, p'_{\perp}) \right\rangle
\nonumber \\
= \delta_{\alpha\delta} \delta_{\beta\gamma}
\delta_{p_{\perp}p'_{\perp}} \delta(E_1 - E_4) \delta(E_2 - E_3)
f_{\alpha} (E_1 + p_{\perp}^2/2m) \left[ 1 - f_{\beta} (E_2 +
p^{'2}_{\perp}/2m) \right].  
\end{eqnarray}
\hspace*{3.6truein}\noindent \hrulefill 
\begin{multicols}{2}
\noindent
After simple but tedious algebra where we 
replace at an intermediate stage $((E_z - E_0)^2 +
\Gamma^2(E_z)/4)^{-2}$ by $4\pi\Gamma^{-3}(E_0) \delta(E_z -
E_0)$, we obtain $S_{LL} = S_{RR} =
-S_{LR} = -S_{RL}$ (thus checking the conservation of current) with
\begin{eqnarray} \label {noise2}
S_{LL} & \equiv & S_{LL} (0) = \frac{2e^2\nu_2}{{\cal{A}}}
\frac{\Gamma_L(E_0)\Gamma_R(E_0)}{\Gamma(E_0)^{3}}
\nonumber \\
& \times &  (v - u + E_F - E_0) \left\{ \Gamma^2_L (E_0) + \Gamma^2_R
(E_0) \right. \nonumber \\
& + & \left. 2 \frac{e{\cal{A}}J_L}{C_L + C_R + C_0} \left[
\Gamma_R(E_0) - \Gamma_L(E_0) \right] \right. \nonumber \\
& + & \left. 2 \left( \frac{e{\cal{A}}J_L}{C_L + C_R +
C_0} \right)^2 \right\}
\end{eqnarray}
for $v_a < v < v^*$ and $S_{LL} = 0$ otherwise. 

With the help of the spectral density of the current noise,
evaluated on the stable branch of the I-V-characteristic, 
we can find the Fano factor. With the interaction energy 
$$\Lambda \equiv \frac{e{\cal{A}}J_L}{C_L + C_R + C_0}, \ \ \
\Gamma_{L,R} \equiv \Gamma_{L,R} (E_0)$$ 
we find 
\begin{eqnarray} \label{Fano2}
F = \frac{\Gamma_L^2 + \Gamma_R^2 + 2 \Lambda \left[\Gamma_R -
\Gamma_L \right] + 2 \Lambda^2}{[\Gamma_L + \Gamma_R]^2} . 
\end{eqnarray}
Eq. (\ref{Fano2}) is the main result of our work and together with
the definitions of $J_L$ and $C_0$ and with the equation
(\ref{self-cons}) for $u(v)$ (the stable charged branch solution) in
an implicit form gives the dependence of the noise power on the
applied voltage $v$. Using Eq. (\ref{Fano2}) for the Fano factor 
and the definition of the asymmetry $\Delta \Gamma = (\Gamma_L -
\Gamma_R)/2$ gives the Fano factor, Eq. (\ref{Fano1}), mentioned in
the introduction.    

The dependence on the Fano factor on the external voltage $v$ is
illustrated in an explicit manner in Fig.~5.  

There are several important points to comment on. 

First, the function $\Lambda(v)$ accommodates the interaction
parameter $C_e$ and represents the contribution of interactions to the
noise power. For $\Lambda = 0$ (no interaction) we reproduce the
free-electron result \cite{ChenTing} $F_0 = (\Gamma_L^2 +
\Gamma_R^2)(\Gamma_L + \Gamma_R)^{-2}$, as already discussed in the
introduction. Note, however, that due to the energy dependence of the
tunneling rates for a symmetric dot, $a_L = a_R$, one has $\Gamma_L <
\Gamma_R$, which implies $1/2 < F_0 < 1$.  

Furthermore, for $v \to v^*$ the denominator $C_L + C_R + C_0(v)$ of
the function $\Lambda (v)$ quite generally diverges as $(v^* -
v)^{-1/2}$, while the function $J_L(v)$ in the numerator stays
finite. Consequently, the Fano factor diverges according to $(v^* -
v)^{-1}$. This is clearly seen in Fig.~5.

For $v=v_a = C_R^{-1}(C_L+C_R)(E_0-E_F)$ the Fano factor can be
calculated in a closed form. We have
$$C_0(v_a) = C_e \frac{\Gamma_L}{\Gamma_L + \Gamma_R},$$
$$J_L(v_a) = - e\nu_2 \frac{\Gamma_L\Gamma_R}{\Gamma_L + \Gamma_R},$$
which gives
$$\Lambda (v_a) = -\frac{C_e\Gamma_L\Gamma_R}{(C_L+C_R)(\Gamma_L +
\Gamma_R) + C_e\Gamma_L} < 0.$$
It is seen that $-\Gamma_R < \Lambda (v_a) < 0$, implying $1/2 <
F(v_a) < 1$. Thus, we have described the transition from 
\begin{figure}
\narrowtext
{\epsfxsize=5.0cm\epsfysize=5.0cm\centerline{\epsfbox{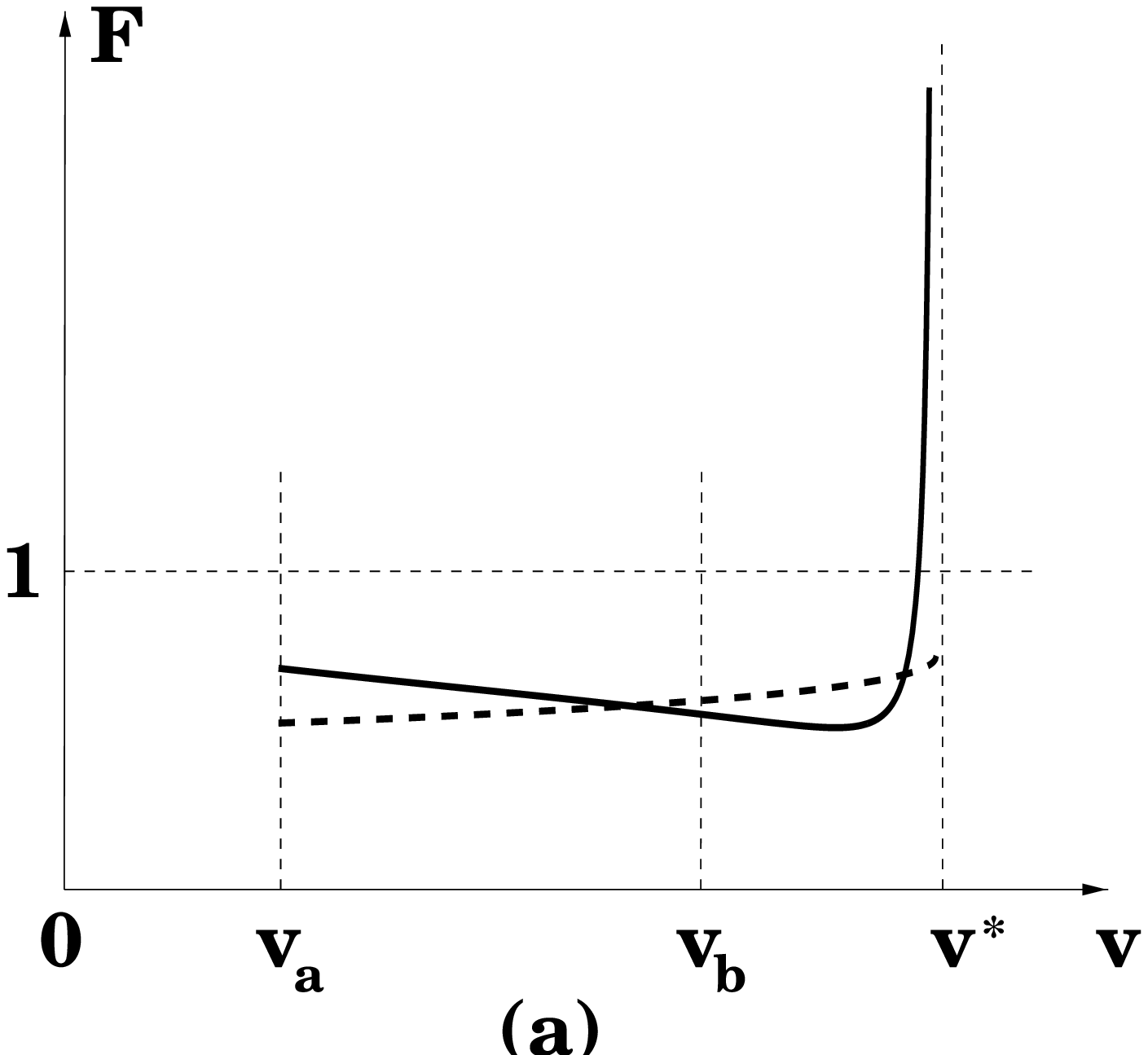}}}
{\epsfxsize=5.0cm\epsfysize=5.0cm\centerline{\epsfbox{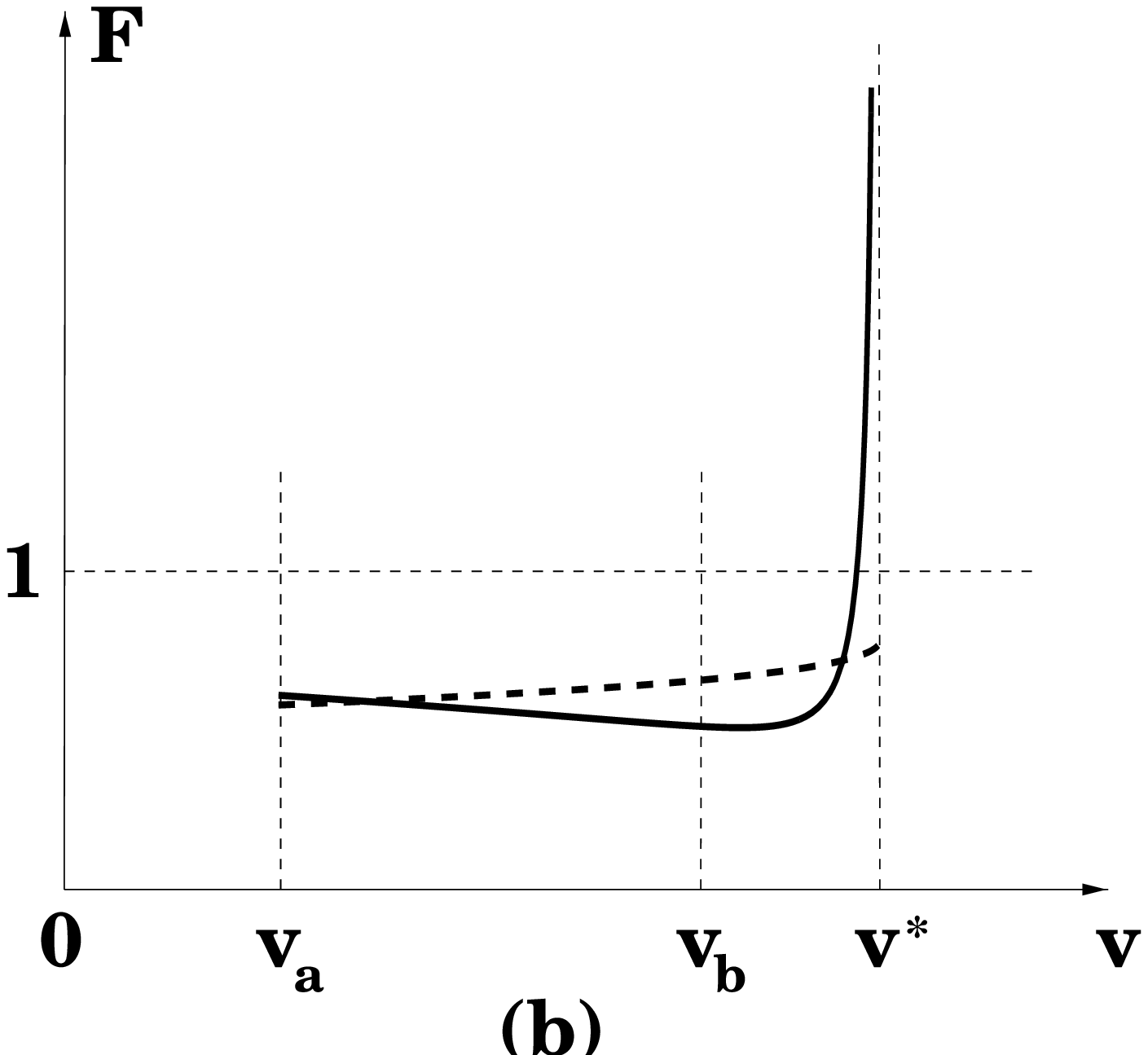}}}
{\epsfxsize=5.0cm\epsfysize=5.0cm\centerline{\epsfbox{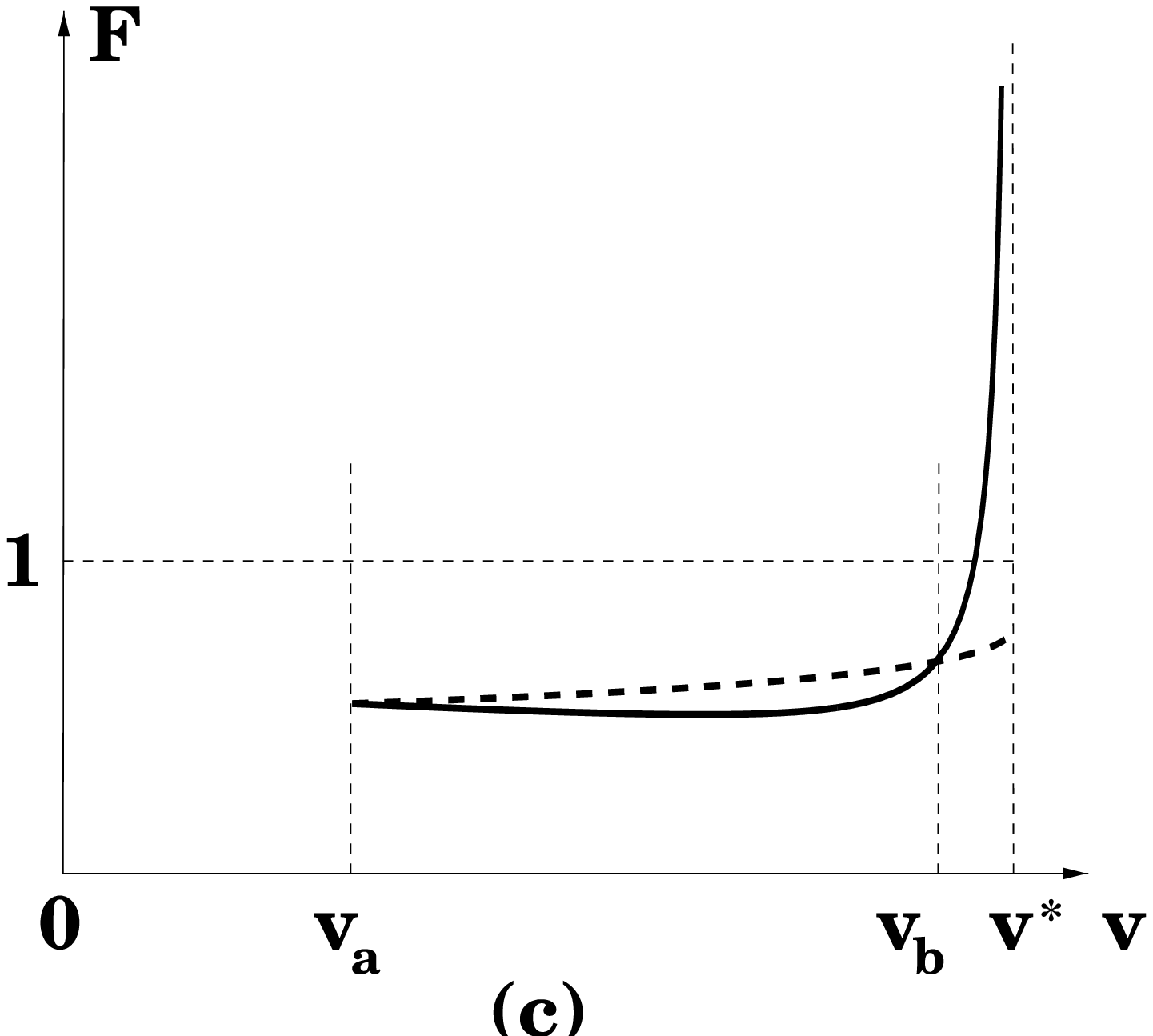}}}
{\epsfxsize=5.0cm\epsfysize=5.0cm\centerline{\epsfbox{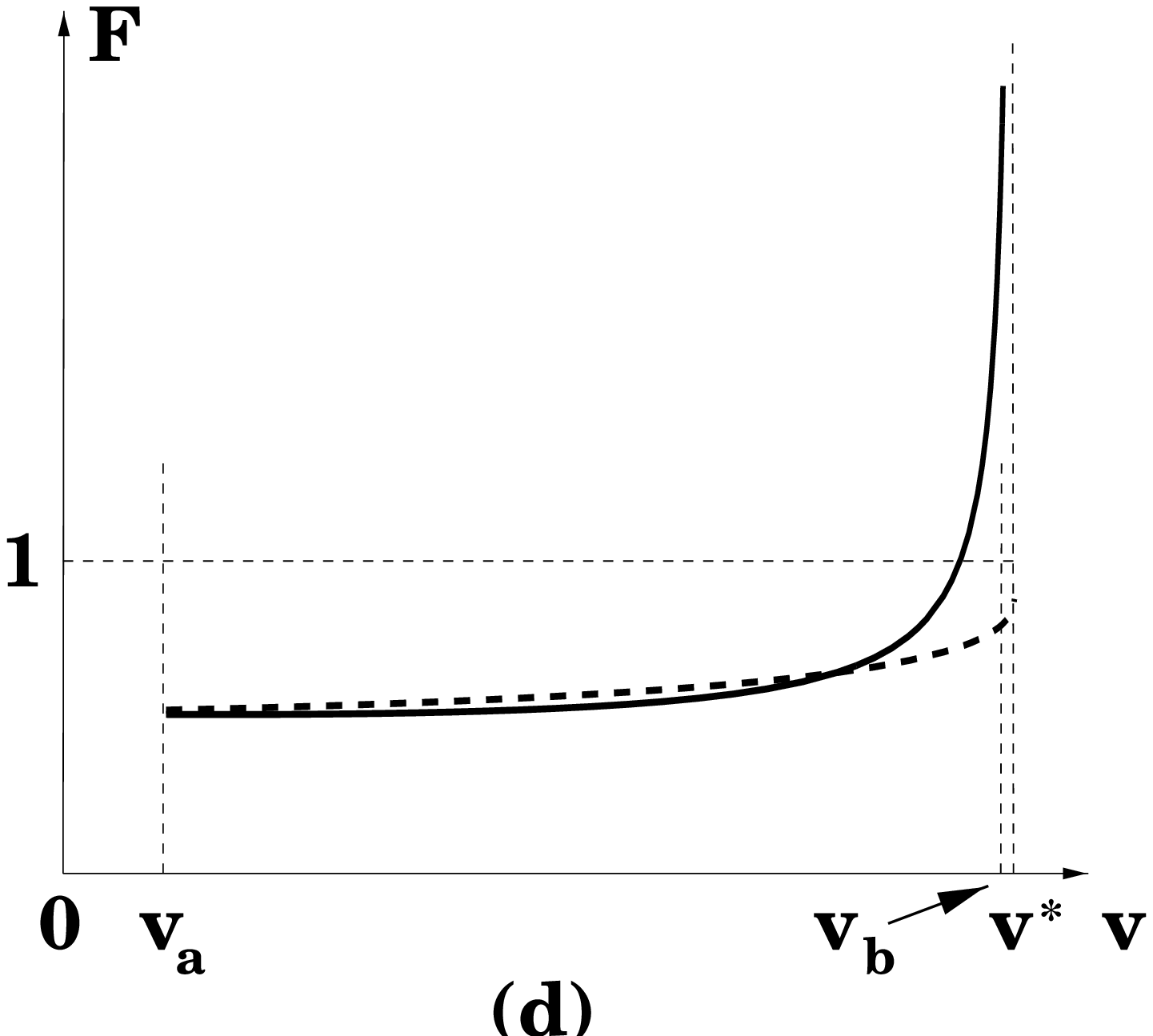}}}

\protect\vspace{0.5cm}

\caption{The full Fano factor $F$ (solid line) and its non-interacting
part $F_0$ (bold dashed line) as a function of 
applied voltage $v$. The parameters are chosen as follows:
$a_L = a_R$; $C_L = C_R$; (a) $C_e = 10C_L$, $E_0 = 3E_F/2$ (this set
corresponds to that of Fig.~2--4); (b) $C_e = 10C_L$, $E_0 = 3E_F$;
(c) $C_e = 4C_L$, $E_0 = 2E_F$; (d) $C_e = C_L$, $E_0 = 3E_F/2$. 
The voltage $v^*$ (in units of $E_0$) is equal to 2.70 (a), 2.26 (b),
2.14 (c), 2.02 (d).} 
\label{fig5}
\end{figure}
\noindent
{\em
sub-Poissonian} to {\em super-Poissonian} shot noise. The range of
super-Poissonian noise ($F(v) > 1$) lies at voltages close to $v^*$;
this range is the  larger the stronger the interaction is (the
parameters $C_e/C_L$ and $C_e/C_R$ grow) and shrinks with increasing
distance of the resonance from the equilibrium Fermi level, i. e. with
increasing $E_F/E_0$. Note that the onset of the super-Poissonian
noise (the voltage where $F[v]=1$) is unrelated to the point of zero
differential resistance ($dj/dv = 0$). Of course, these two points may
lie quite close to each other. In addition, the quantity $J_L$ changes
sign from negative to positive when the voltage approaches $v^*$. This
implies that $\Lambda$ also changes sign from negative to positive at
the same voltage. 

Finally, with respect to the role of interactions, two scenarios 
emerge: To avoid cumbersome expressions, we perform the analysis
for the symmetric case $a_L = a_R$, $C_L = C_R$. Then for ``weak
interactions'', 
$$C_e < \tilde C \equiv 4C_L \left( \frac{E_0}{E_F} - 1 \right),$$ 
we have $F(v_a) < F_0(v_a)$. Thus, the curves $F(v)$ and $F_a(v)$
cross each other only once, when $\Lambda (v) = 0$. This means that
interactions suppress the shot noise for ``low'' voltages and enhance
it for ``high'' voltages, eventually driving the noise power to
super-Poissonian values for voltages close to $v^*$. For ``strong
interactions'', $C_e > \tilde C$, we obtain $F(v_a) > F_0 (v_a)$, and
the curves cross each other twice. In this case, interactions first 
enhance the shot noise, then suppress it, and then enhance it
again. In the limiting case $C_e \gg C_L$ one has $F(v_a) = 1$; then
the shot noise starts from the Poissonian value. Fig.~5a and Fig.~5b
represent the ``strong interaction'' scenario. Fig.~5d illustrates the
case of ``weak interactions'', while Fig.~5c shows the marginal case
$C_e = \tilde C$, where one has $F(v_a) = F_0 (v_a)$.   

\section{Conclusions}

We now discuss the assumptions we made in the course of the derivation
of our result, Eq. (\ref{Fano1}). First, in order to calculate the
shot noise, we linearized the charge and current operators in the
fluctuations of the band bottom. Close to the voltage $v^*$, at the
linear stability threshold, where the solutions corresponding to the
charged well vanish, these fluctuations become large, and the
linearization is not justified any more. Since we evaluate only the
zero-frequency response, the quantity $\large \hat u^2(t)$ is not
amenable to our treatment, and the limits of this approximation can
not be determined. We emphasize, that in the crossover region from
sub- to super-Poissonian noise our approach is well justified. Thus,
our discussion can be used to determine the range of voltages in which
super-Poissonian noise can be observed.  

To determine the effect of interactions on the Fano factor, we 
have neglected the finite width of the resonant level in the 
evaluation of $\Lambda$. For comparison with experiment the broadening
of the resonant level (finite $\Gamma$) should be taken into account. 
This would smear all the curves; in particular, the noise power at $v
= v^*$ would acquire a finite value. Moreover, if the tunneling rate
$\Gamma$ exceeds the hysteresis interval $v^* - v_b$, an extended
range of differential resistance will be observed instead of the
hysteretic behavior. We believe that is the case in the experiments of
Ref. \onlinecite{Iannaccone}. Note also that in our treatment the
interactions make the Fano factor discontinuous at $v = v_a$, where
it jumps from the non-interacting value\cite{ChenTing} $F_0(v_a)$ to
the renormalized one $F(v_a)$. This discontinuity is also healed by
the introduction of the finite width of the resonant level. 

In our work we have treated a single resonant level. It is clear that
wells which permit to fill a number of subbands can be expected to
exhibit even richer behavior than indicated in this work. Since
interactions play a dominant role in the wide area resonant double
barrier structure considered here, it might be interesting to analyze
the noise power of a single discrete level in a quantum dot under high
bias and ask if in fact a similar behavior results. It would also be
interesting to investigate the effect of potential fluctuations not
only in the well but also in the accumulation and depletion layers in
the cathode and anode. In the terminology of
Ref. \onlinecite{Davies2} we have in such a model internal degrees of
the well which couple to the cathode. This would modify the current
injection into the well (depending on the charge state) and possibly
provides a mechanism which can lead to Fano factors smaller than $F =
1/2$.     

In conclusion, we investigated current-voltage characteristics and
noise power in resonant tunneling quantum wells in the nonlinear
regime. We treated the low-transparency case, when the total width 
$\Gamma$, due to tunneling, is much lower than all other relevant 
energy scales. (For the evaluation of the interaction effects, we 
have taken the transmission coefficient as delta-function in
energy). The interaction is taken into account via the charge
accumulated in the well due to capacitive coupling to the
reservoirs. In this regime, we discovered that interactions
dramatically affect transport properties of the quantum well. They
produce a range of voltages ($v_b < v < v^*$) where the well can be in
one of two stable states, which correspond to a charged well and an
uncharged well, respectively. For the current-voltage characteristics,
this leads to a hysteretic behavior, as shown in Fig.~4. Interaction
effects are even more pronounced in the noise power: At low voltages
interactions can either suppress or enhance the noise power below or
above the shot noise level predicted by a free-electron theory,
and at large voltages the noise power diverges as $(v^* - v)^{-1}$ 
for $v \to v^*$, and becomes super-Poissonian close to $v^*$. 
Both the noise suppression below the non-interacting
value\cite{Brown,Mendez} and the strong enhancement of the noise to
super-Poissonian values\cite{Iannaccone,Mendez} have been observed
experimentally. We stress that the range of super-Poissonian noise is
generally different from the negative differential resistance range,
and may extend well outside the hysteresis range (see the example
shown in Fig.~5d). We believe that our results provide an explanation
of the role of interactions on the shot noise in resonant wells.  

Clearly, the interaction effects in the noise power which we have
discussed are very interesting. While the search for interaction
effects in mesoscopic physics represents a huge effort, there are up
to now, at best very few clearly identified signatures for which both
experiments and theory agree. A notable exception are of course the
Coulomb blockade effects. We emphasize that the effects discussed
here, occur in a state which is a far from equilibrium transport
state. Resonant tunneling systems have the important advantage that
they represent, at least conceptually, a very simple system and thus
provide a unique testing ground for both theories and experiments.  

\section*{Acknowledgements}

We thank E.~E.~Mendez and V.~V.~Kuznetsov who sent us their results
prior to publication. The work was supported by the Swiss National
Science Foundation.

\end{multicols}

\begin{references}
\bibitem{note1} The capacitance $C_{0}[u]$ plays here the same role 
as the polarization function of the transport state defined as 
$\partial \langle Q \rangle = - e \Pi[u] \partial u$. We have 
$C_{0}[u] \equiv e^{2} \Pi[u].$
\end{references}
\end{document}